\documentclass[graybox]{svmult}

\usepackage[numbers, compress]{natbib}
\usepackage{mathptmx}       
\usepackage{helvet}         
\usepackage{courier}        
\usepackage{type1cm}        
\usepackage{makeidx}         
\usepackage{graphicx}        
\usepackage{multicol}        
\usepackage[bottom]{footmisc}
\usepackage{amsmath}
\usepackage{mathtools}

\makeindex             

\begin{document}

\title*{Scheduling Data-Intensive Workloads in Large-Scale Distributed Systems: Trends and Challenges}
\titlerunning{Scheduling Data-Intensive Workloads in Large-Scale Distributed Systems}  
\author{Georgios L. Stavrinides and Helen D. Karatza}
\institute{Georgios L. Stavrinides \at Department of Informatics, Aristotle University of Thessaloniki, 54124 Thessaloniki, Greece, \email{gstavrin@csd.auth.gr}
\and Helen D. Karatza \at Department of Informatics, Aristotle University of Thessaloniki, 54124 Thessaloniki, Greece, \email{karatza@csd.auth.gr}}
%
%
\maketitle

\abstract*{With the explosive growth of big data, workloads tend to get more complex and computationally demanding. Such applications are processed on distributed interconnected resources that are becoming larger in scale and computational capacity. Data-intensive applications may have different degrees of parallelism and must effectively exploit data locality. Furthermore, they may impose several Quality of Service requirements, such as time constraints and resilience against failures, as well as other objectives, like energy efficiency. These features of the workloads, as well as the inherent characteristics of the computing resources required to process them, present major challenges that require the employment of effective scheduling techniques. In this chapter, a classification of data-intensive workloads is proposed and an overview of the most commonly used approaches for their scheduling in large-scale distributed systems is given. We present novel strategies that have been proposed in the literature and shed light on open challenges and future directions.}

\abstract{With the explosive growth of big data, workloads tend to get more complex and computationally demanding. Such applications are processed on distributed interconnected resources that are becoming larger in scale and computational capacity. Data-intensive applications may have different degrees of parallelism and must effectively exploit data locality. Furthermore, they may impose several Quality of Service requirements, such as time constraints and resilience against failures, as well as other objectives, like energy efficiency. These features of the workloads, as well as the inherent characteristics of the computing resources required to process them, present major challenges that require the employment of effective scheduling techniques. In this chapter, a classification of data-intensive workloads is proposed and an overview of the most commonly used approaches for their scheduling in large-scale distributed systems is given. We present novel strategies that have been proposed in the literature and shed light on open challenges and future directions.}


\section{Introduction}
\label{sec:intro}
The ever-increasing momentum of the Internet of Things, the rapid pace of technological advances in mobile devices and cloud computing, as well as the explosive growth of social media, produce an overwhelming flow of data of unprecedented volume and variety at a record rate. Such data are commonly referred to as \emph{big data} and are characterized by the following attributes: (a) volume, i.e. they consist of very large datasets, (b) variety, i.e. they comprise diverse structured and unstructured data of various types  and (c) velocity, i.e. the data are generated and streamed at staggering speeds~\cite{Russom2011, Hashem2015}. Computationally intensive applications are employed in a wide spectrum of domains such as healthcare, science, engineering, business and finance, in order to unleash the power of big data, extract useful knowledge and gain valuable insights.~\cite{Talia2013}.   

The advent of big data has called for a paradigm shift in the computer architecture, and consequently the applications, required  for their effective processing. Data-intensive applications are typically processed on interconnected computing resources that are geographically distributed, encompass various heterogeneous components, utilize virtualization, feature multi-tenancy and are able to scale up in the foreseeable future. Computer clusters, computational grids and clouds are examples of such platforms~\cite{Foster2008}. Furthermore, novel hybrid approaches have emerged, such as \emph{fog computing}, which extends the cloud computing paradigm by bringing data processing at computational resources at the edge of the network, closer to where the data are generated, while sending selected data to the cloud for historical analysis and long-term storage~\cite{Cisco2015, Bonomi2014}.  

Data-intensive applications may have different degrees of parallelism and must effectively exploit data locality. Furthermore, they may also impose several Quality of Service (QoS) requirements, such as time constraints and resilience against failures, as well as other objectives, like energy efficiency. These features of the workloads operating on big data, as well as the characteristics of the computing resources required to process them, present major challenges that require the employment of effective \emph{scheduling algorithms}. Due to their inherent complexity, the performance of such algorithms is usually evaluated by simulation, rather than by analytical methods. Analytical modeling is difficult and often requires several simplifying assumptions that may have an unpredictable impact on the results~\cite{Stavrinides2017c}.

This chapter is organized as follows: Sect.~\ref{sec:schedulingProblem} gives a definition of the scheduling problem in large-scale distributed systems, as well as some of the most important scheduling objectives. In Sect.~\ref{sec:dataIntensiveApplications}, a classification of data-intensive workloads is proposed, according to their degree of parallelism. An overview of the most widely used strategies for the scheduling of each class of data-intensive applications in large-scale distributed systems is given. Sect.~\ref{sec:challenges} presents some of the major challenges of data-intensive workload scheduling, covering topics such as  data locality awareness, timeliness, fault tolerance and energy efficiency. Furthermore, novel strategies that have been proposed in the literature are presented in Sect.~\ref{sec:novelStrategies}. Finally, Sect.~\ref{sec:conclusions} concludes this chapter, shedding light on open challenges and future research directions.

\section{Scheduling Problem}
\label{sec:schedulingProblem}
In its general form, the scheduling problem in large-scale distributed systems concerns the mapping of a set of application tasks $V=\{ n_{1}, n_{2},..., n_{N}\}$ to a set of processors $P=\{ p_{1}, p_{2},..., p_{Q} \}$, in order to complete all tasks under the specified constraints (e.g. complete each task within its deadline)~\cite{Buttazzo2011, Kolodziej2012}. In this general form, the scheduling problem has been shown to be NP-complete~\cite{Garey1979}.

\subsection{Scheduling Objectives}
\label{subsec:objectives}
\begin{figure}[t!]
\centering
\includegraphics[scale=0.75]{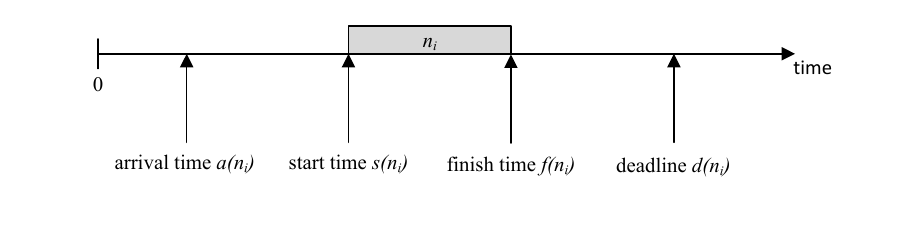}
\caption{Typical parameters that characterize a task of an application submitted for execution in a large-scale distributed system.}
\label{fig:taskTimes}
\end{figure}

Some of the parameters that characterize a task $n_{i} \in V$ are shown in Fig.~\ref{fig:taskTimes}. These parameters are:
\begin{itemize}
\item \emph{arrival time} $a(n_{i})$: it is the time at which the task arrives at the system.
\item \emph{start time} $s(n_{i})$: it is the time at which the task starts its execution.
\item \emph{finish time} $f(n_{i})$: it is the time at which the task finishes its execution.
\item \emph{deadline} $d(n_{i})$: it is the time before which the task should finish its execution.
\end{itemize}

Based on the above parameters, some of the most commonly used scheduling objectives in large-scale distributed systems are:
\begin{description}
\item[\textbf{(a)}] To minimize the \emph{average response time} $\overline{R}$ of the tasks $n_{i} \in V$, where $\overline{R}$ is given by:
\begin{equation}
\overline{R}= \frac{1}{N} \sum_{n_{i} \in V} R(n_{i})
\end{equation}  
where $R(n_{i})=f(n_{i}) - a(n_{i})$ and $N$ is the number of tasks in $V$.

\item[\textbf{(b)}] To minimize the \emph{makespan} (i.e. total execution time) $M$ of the tasks $n_{i} \in V$, where $M$ is defined as:
\begin{equation}
M= \max_{n_{i} \in V} { \left \{  f(n_{i}) \right \} } - \min_{n_{i} \in V} { \left \{ s(n_{i}) \right \} }
\end{equation}

\item[\textbf{(c)}] To maximize the \emph{task guarantee ratio} $TGR$ of the tasks $n_{i} \in V$, where $TGR$ is given by:
\begin{equation}
TGR= \frac{1}{N} \sum_{n_{i} \in V} guar(n_{i})
\end{equation}
where
\begin{equation}
guar(n_{i})=
\begin{dcases}
1 & \text{if } f(n_{i}) \leq d(n_{i})\\
0 & \text{otherwise}
\end{dcases}
\end{equation}
 
\item[\textbf{(d)}] To minimize the \emph{average tardiness} $\overline{T}$ of the tasks $n_{i} \in V$, where $\overline{T}$ is defined as:
\begin{equation}
\overline{T}= \frac{1}{N} \sum_{n_{i} \in V} T(n_{i})
\end{equation}
where 
\begin{equation}
T(n_{i})=
\begin{dcases}
f(n_{i}) - d(n_{i}) & \text{if } f(n_{i}) > d(n_{i})\\
0 & \text{otherwise}
\end{dcases}
\end{equation}
\end{description}

\section{Data-Intensive Workloads in Large-Scale Distributed Systems}
\label{sec:dataIntensiveApplications}
The data-intensive applications scheduled for execution in large-scale distributed systems, typically consist of numerous component tasks. At the one end of the spectrum, the tasks require  frequent communication with each other during their execution. At the other end of the spectrum, the component tasks do not require any communication and are completely independent. Between these two ends,  is the case where communication is required between the component tasks of an application, but only before or after their execution. Consequently, data-intensive workloads in large-scale distributed systems can be classified into the following categories:
\begin{itemize}
\item \emph{fine-grained parallel applications},
\item \emph{coarse-grained parallel applications} and
\item \emph{embarrassingly parallel applications}.
\end{itemize}
In the following paragraphs, each class of data-intensive applications is presented in more detail and their corresponding, most widely used scheduling heuristics are analyzed.

\subsection{Fine-Grained Parallel Applications}
\label{subsec:gangs}
\begin{figure}[t!]
\centering
\includegraphics[scale=1.5]{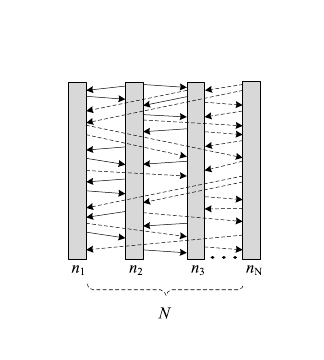}
\caption{An example of a fine-grained parallel application. The frequently communicating tasks of the application form a gang of $N$ parallel tasks. The communication between the tasks is depicted with arrows.}
\label{fig:gang}
\end{figure}

An application features \emph{fine-grained parallelism} when it consists of frequently communicating parallel tasks. A proven and effective way to schedule such applications is \emph{gang scheduling}. According to this approach, the parallel tasks of an application form a \emph{gang} and are scheduled and executed simultaneously on different processors. Hence, all of the tasks of the application  start execution at the same time. This way, the risk of a task waiting to communicate with another task that is currently not running is avoided. The task with the largest execution time determines the execution time of the gang. An example of a gang with $N$ parallel tasks is shown in Fig.~\ref{fig:gang}.

\begin{figure}[t!]
\centering
\includegraphics[scale=0.95]{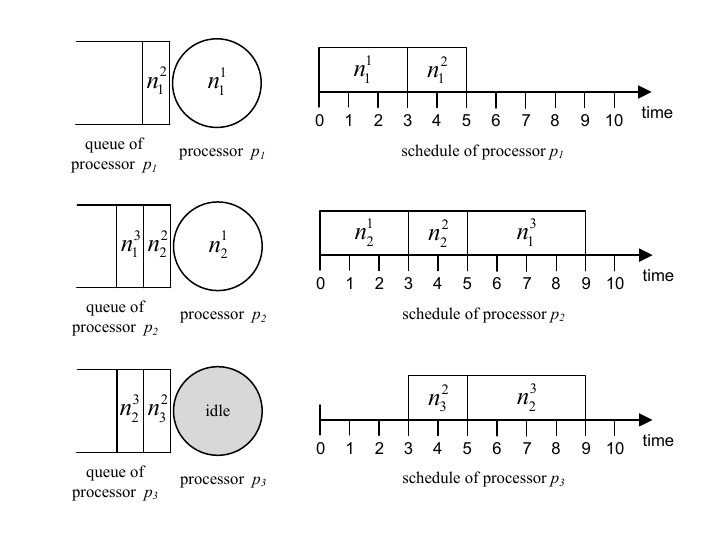}
\caption{Example of gang scheduling in a system with three processors $p_{1}$, $p_{2}$ and $p_{3}$. The first gang consists of the tasks $n_{1}^{1}$ and $n_{2}^{1}$, scheduled on processors $p_{1}$ and $p_{2}$, respectively. The second gang consists of the tasks $n_{1}^{2}$, $n_{2}^{2}$ and $n_{3}^{2}$, scheduled on processors $p_{1}$, $p_{2}$ and $p_{3}$, respectively. The third gang consists of the tasks $n_{1}^{3}$ and $n_{2}^{3}$, scheduled on processors $p_{2}$ and $p_{3}$, respectively. It can be observed that the processor $p_{3}$ remains idle during the execution of the tasks $n_{1}^{1}$ and $n_{2}^{1}$ of the first gang. This is due to the fact that the task $n_{3}^{2}$ at the head of its queue cannot start execution, because according to the gang scheduling technique, it must start execution at the same time as the other tasks of its gang, $n_{1}^{2}$ and $n_{2}^{2}$, which are scheduled on the other processors that are currently busy.}
\label{fig:gangScheduling}
\end{figure}

Consequently, gang scheduling facilitates the synchronization between the component tasks of a fine-grained parallel application. Without this technique, the synchronization of the component tasks would require more context switches and thus additional overhead. On the other hand, in order to utilize gang scheduling, the number of available processors must be greater than or equal to the number of parallel tasks of the application. Furthermore, due to the requirement that all of the tasks of a gang must start execution at the same time, there may be times at which some of the processors are idle, even with tasks waiting in their respective queues. Specifically, a task at the head of the queue of an idle processor may be waiting for the other tasks of its gang, which may not be able to start execution at the particular time instant~\cite{Stavrinides2016a}. This situation is depicted in Fig.~\ref{fig:gangScheduling}.

\subsubsection{Gang Scheduling Policies}
\label{subsubsec:gangScheduling}
The two most widely used gang scheduling policies are the \emph{Adapted First Come First Served (AFCFS)} and \emph{Largest Gang First Served (LGFS)} strategies.

\paragraph{Adapted First Come First Served (AFCFS)} 
\label{para:afcfs}
This method is an adapted version of the First Come First Served (FCFS) scheduling heuristic,  according to which the gang that arrived first, has the highest priority for execution. A gang starts execution when its tasks are at the head of their assigned queues and the respective processors are idle. When there are not enough idle processors for a gang with a large number of parallel tasks waiting at the front of their assigned queues, a smaller gang with tasks waiting behind those of the larger gang can start execution. This technique is also referred to as \emph{backfilling}~\cite{Karatza2008}.

The major drawback of this scheduling policy is that it tends to favor smaller gangs, which leads to greater response times for larger gangs. In order to overcome this issue, various techniques have been proposed in the literature, such as the employment of a \emph{bypass count} parameter~\cite{Manickam2012} and the utilization of \emph{task migrations}~\cite{Papazachos2009}. The first method, counts for each gang the number of gangs that bypassed it, due to an insufficient number of idle processors. When the bypass count of a gang reaches a specified threshold, it gets the highest priority for execution. According to the second method, the tasks of a gang are candidate for migration only if at least one of them is at the head of its assigned queue and the respective processor is idle. The tasks that are migrated, are placed at the head of their newly assigned queues. In order to avoid the starvation of the other tasks, there is a limit on the number of migrated tasks a queue can accept.

\paragraph{Largest Gang First Served (LGFS)}  
\label{para:lgfs}
 According to this scheduling strategy, the tasks in the processor queues are sorted in descending order of gang size (i.e. number of tasks) of their respective gang. Thus, tasks that belong to larger gangs have higher priority than tasks that belong to smaller gangs. Whenever a processor becomes idle, the scheduler searches the queues starting from the head of each queue and the first gang with tasks that can start execution occupies the processors~\cite{Karatza2014}. Clearly, this strategy tends to favor applications with a high degree of parallelism (i.e. large gangs), at the expense of smaller gangs. However, this is sometimes desirable and may lead to a better system performance, compared to the AFCFS policy.

\subsection{Coarse-Grained Parallel Applications}
\label{subsec:dags}
In case an application exhibits \emph{coarse-grained parallelism}, its component tasks do not require any communication with each other during processing, but only before or after their execution. That is, the component tasks have precedence constraints among them, in such a way that the output data of a task  are used as input by other tasks. A component task can only start execution when its predecessor tasks have completed. A task without any parent tasks is called an \emph{entry task}, whereas  a task without any child tasks is called an \emph{exit task}. 

Such an application is often called a \emph{workflow application} and  can be represented by a \emph{Directed Acyclic Graph (DAG)}  or \emph{task graph}, $G=(V,E)$, where $V$ and $E$ are the sets of the nodes and the edges of the graph, respectively~\cite{Stavrinides2011a, Stavrinides2014a, Stavrinides2014b}. Each node represents a component task, whereas a directed edge between two tasks represents the data that must be transmitted from the first task to the other. Each node has a weight that represents the computational cost of its corresponding task. Each edge between two tasks has a weight that denotes the communication cost that is incurred when transferring data from the first task to the other. 

\begin{figure}[t]
\centering
\includegraphics[scale=0.95]{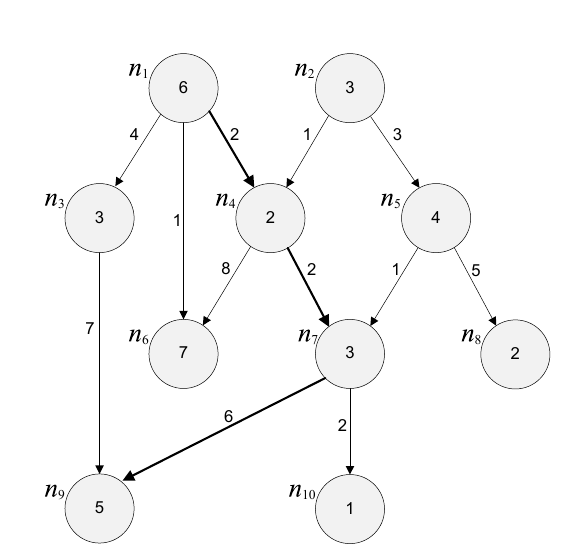}
\caption{An example of a coarse-grained parallel application (workflow application), represented as a Directed Acyclic Graph (DAG). The number in each node denotes the computational cost of the represented task. The number on each edge denotes the communication cost between the two tasks that it connects. The critical path of the DAG is depicted with thick arrows.}
\label{fig:dag}
\end{figure}

The \emph{level} of a task in the graph is equal to the length of the longest path from the particular task to an exit task in the graph. The length of a path is the sum of the computational and  communication costs of all of the nodes and edges, respectively, along the path. The \emph{critical path} of the graph is the longest path from an entry task to an exit task in the graph. An example of a workflow application is illustrated in Fig.~\ref{fig:dag}.

\subsubsection{Workflow Scheduling Approaches}
\label{subsubsec:workflowScheduling}
Workflow applications require a scheduling strategy that should take into account the precedence constraints among their component tasks. The workflow scheduling heuristics are classified into the following general categories:
\begin{itemize}
\item \emph{list scheduling algorithms,}
\item \emph{clustering algorithms,}
\item \emph{task duplication algorithms} and
\item \emph{guided random search algorithms}.
\end{itemize}
These techniques are analyzed in the following paragraphs.

\paragraph{List Scheduling Algorithms} 
\label{para:list}
A list scheduling algorithm consists of two phases: (a) a \emph{task selection phase} and (b) a \emph{processor selection phase}.  In the first phase, the tasks are prioritized based on specific criteria and are arranged in a list according to their priority. The task with the highest priority is selected first for scheduling. During the second phase, the selected task is scheduled to the processor that minimizes  a specific cost function, such as the estimated start time of the task~\cite{Stavrinides2017a}. List scheduling algorithms are the most commonly used among the workflow scheduling heuristics, because they are generally simpler, more practical, easier to implement and they usually outperform other techniques, incurring less scheduling overhead~\cite{Topcuoglu2002}.

One of the simplest list scheduling policies is the \emph{Highest Level First (HLF)}~\cite{Adam1974}. According to this method, the task prioritization phase is based on the level of each task. In the processor selection phase, the selected task is scheduled to the processor that can provide it with the earliest start time. An improved version of the HLF strategy is the \emph{Insertion Scheduling Heuristic (ISH)}~\cite{Kruatrachue1987} and it is based on the observation that idle time slots may form in the schedule of a processor (schedule gaps), due to the data dependencies among the tasks. The task selection phase of this technique is based on HLF. However, during the processor selection phase, a task may be inserted into a schedule gap, as long as it does not delay the execution of the succeeding task in the schedule and provided that it cannot start earlier on any other processor. An alternative version of  ISH, adapted for heterogeneous systems, is the \emph{Heterogeneous Earliest Finish Time (HEFT)} policy~\cite{Topcuoglu2002}. According to this approach, for the calculation of the level of each task, the average computational and communication costs of the tasks and edges, respectively, are used.

\paragraph{Clustering Algorithms}
\label{para:clustering}
The main idea of clustering algorithms is the minimization of the communication cost between the tasks of a DAG, by grouping heavily communicating tasks into the same cluster and assigning all of the tasks in the cluster to the same processor. A clustering algorithm is an iterative process. At first, each task is an independent cluster. At each iteration, previous clusters are refined by merging some of them, according to specific criteria. At the end of the process, a cluster merging step is needed, so that the number of clusters is equal to the number of processors. Subsequently, a cluster mapping step is required, in order to map each cluster to a processor. Finally, a task ordering step is performed, in order to determine the execution order of tasks on each processor~\cite{Jiang2011}.

One of the most popular clustering techniques is the \emph{Dominant Sequence Clustering (DSC)} algorithm~\cite{Yang1994}. This method is based on the observation that the makespan of a DAG is determined by the longest path in the scheduled task graph and not by its critical path, which is calculated before the scheduling of the tasks of the DAG. The longest path in the scheduled DAG is called the \emph{dominant sequence (DS)}.  According to the DSC algorithm, the tasks in a DAG are clustered in such a way, so that the dominant sequence of the graph is minimized.

\paragraph{Task Duplication Algorithms}
\label{para:duplication}
 In this category of workflow scheduling heuristics, the main concept is to utilize idle resource time by duplicating predecessor tasks in a DAG, so that the makespan of the particular DAG is minimized. The various duplication-based algorithms  differentiate with each other, according to the criteria used for the selection of the tasks for duplication.  One of the major drawbacks of task duplication algorithms, is that they usually have higher complexity than the other DAG scheduling techniques.
 
 One of the most well-known duplication algorithms is the \emph{Duplication Scheduling Heuristic (DSH)}~\cite{Kruatrachue1987}. According to this approach, the tasks in a DAG are prioritized according to their level. At each scheduling step, the task with the highest level is selected and is allocated to the processor that can provide it with the earliest start time. In order to calculate the earliest possible start time of  the selected task on each processor, first its start time is calculated without duplication of any predecessor tasks. Subsequently, the \emph{duplication time slot} is determined, which is the time period between the finish time of the last scheduled task on the particular processor and the start time of the currently examined task. The algorithm then tries to duplicate the predecessors of the task into the duplication time slot in a recursive manner, starting from the parent task from which the data arrives the latest, until either the slot cannot accommodate other predecessor tasks or the start time of the examined task is not improved.

\paragraph{Guided Random Search Algorithms}
\label{para:random}
A guided random search algorithm is an iterative process of finding the best schedule for a DAG, based on specific criteria. At each step, the previously generated schedule is improved, by utilizing random parameters for the generation of the new schedule. This iterative process terminates according to a predefined condition. These algorithms, even though they generally generate schedules of good quality, however, they incur a much higher scheduling overhead than the other workflow scheduling methods. The most commonly used algorithms of this category are \emph{genetic algorithms}, according to which each new schedule is generated by applying evolutionary techniques from nature, known as \emph{fitness functions}~\cite{Gkoutioudi2012}.

\emph{Simulated Annealing (SA)} is another example of  a guided random search meta-heuristic.  This technique emulates the physical process of annealing in metallurgy, which involves the heating and the controlled, slow cooling of metals, in order to form a crystallized structure without any defects~\cite{Moschakis2015}.  In SA, a temperature variable is used in order to simulate this heating process. Initially, it is set at a high value and as the algorithm runs, it is allowed to slowly cool down. While the value of the temperature variable is high, the algorithm is allowed to accept solutions that are worse than the current one, with higher frequency. As the value of the temperature variable is decreased, so is the chance of accepting worse solutions. Therefore, the algorithm gradually focuses on an area of the search space in which hopefully a near-optimal solution can be found.

\subsection{Embarrassingly Parallel Applications}
\label{subsec:bots}
\begin{figure}[t]
\centering
\includegraphics[scale=0.95]{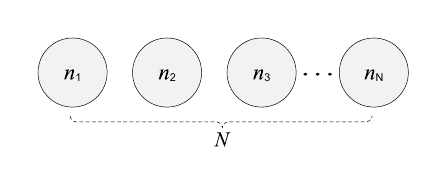}
\caption{An embarrassingly parallel application, consisting of $N$ independent parallel tasks. Such applications are commonly referred to as Bag-of-Tasks (BoT) applications.}
\label{fig:bot}
\end{figure}

An application is regarded as \emph{embarrassingly parallel} when its component tasks are independent, do not communicate with each other and can be executed in any order. Due to these characteristics, such applications are also called \emph{Bag-of-Tasks (BoT)} applications. Due to the independence between their tasks, BoT applications are well suited for execution on widely distributed resources, such as computational grids, where communication can become a bottleneck for more tightly-coupled parallel applications, such as gangs and DAGs~\cite{Weng2005, Stavrinides2017d, Stavrinides2017e}. An example of a BoT application is depicted in Fig.~\ref{fig:bot}.

\subsubsection{Scheduling BoT Applications}
\label{subsubsec:botScheduling}
The most widely used strategies for scheduling BoT applications are: (a) \emph{Min-Min}, (b) \emph{Max-Min} and (c) \emph{Sufferage}. All of these policies focus on minimizing the makespan of the scheduled BoT application.

\paragraph{Min-Min}
\label{para:minmin}
This heuristic is an iterative process, consisting of two steps. In the first step, the \emph{minimum completion time (MCT)} of each unassigned task is calculated, over all of the processors in the system. In the second step, the task with the minimum MCT is assigned to the corresponding processor. At each iteration of the algorithm, the MCT of each unassigned task is determined taking into account the current load of the processors, as resulted by the scheduling decision of the previous iteration~\cite{Weng2005}.     

\paragraph{Max-Min}
\label{para:maxmin}
This strategy differs from the Min-Min policy, in that the task with the maximum (instead of the minimum) MCT is assigned to the corresponding processor in the second step of the scheduling process. Consequently, in cases where the application consists of a large number of tasks with small execution times and a few tasks with large execution times, the Max-Min heuristic is likely to give a smaller makespan than the Min-Min algorithm, since it schedules the tasks with larger execution times at earlier iterations~\cite{Tabak2014}.

\paragraph{Sufferage}
\label{para:suffereage}
This algorithm is a two-step iterative process, like the Min-Min and Max-Min heuristics. However, in this case, in addition to the MCT of each task, its second MCT is also calculated during the first step of the process. Subsequently, the \emph{sufferage value} of each task is determined, by subtracting its MCT from its second MCT. In the second step, the task with the largest sufferage value is assigned to the processor that can provide it with the MCT. That is, this heuristic is based on the idea that the highest priority for scheduling should be given to the task that would suffer the most (in terms of completion time) if it is not assigned to the processor that can provide it with the MCT~\cite{Maheswaran1999}.

\section{Major Challenges}
\label{sec:challenges}
In addition to the challenges imposed by their degree of parallelism, data-intensive applications in large-scale distributed systems must also effectively exploit data locality. Furthermore, they may have various QoS requirements, such as timeliness and fault tolerance, as well as other objectives, like energy efficiency. These requirements are usually specified in a \emph{Service Level Agreement (SLA)}, which is a contract between the user that submits the application for execution and the provider of the infrastructure that the application is executed on. In the following paragraphs, representative examples for each case are given.

\subsection{Data Locality}
\label{subsec:dataLocality}
The most important aspect of scheduling data-intensive applications in large-scale distributed systems is the effective exploitation of data locality. That is, the tasks that operate on big data should be allocated to computational resources that are as near as possible to where the data reside, so that the communication cost incurred by transferring for processing vast amounts of data  from remote resources is minimized.  

\subsubsection{MapReduce \& Hadoop}
\label{subsubsec:mapReduce}
The MapReduce programming paradigm has been proposed by Google~\cite{Dean2008} and facilitates the massively parallel processing of large volumes of data. It is inspired by the map and reduce functions commonly used in functional programming. A MapReduce application consists of two types of tasks: (a) a \emph{map task} and (b) a \emph{reduce task}. A map task takes a set of data and converts it into another set of data, where individual elements are broken down into tuples (i.e. key/value pairs). Parallel map tasks can process different chunks of data. A reduce task takes as input the output from map tasks and combines those data tuples into a smaller set of tuples, in order to produce the final result. A reduce task is always performed after the map tasks. In case a MapReduce application has only map tasks, it is considered an embarrassingly parallel application. In case an application has one or more reduce tasks, it is considered a coarse-grained parallel application. In the latter case, multiple reduce tasks can be employed in order to enhance the parallelism of the application~\cite{Ekanayake2010}. 

\begin{figure}[h]
\centering
\includegraphics[scale=0.90]{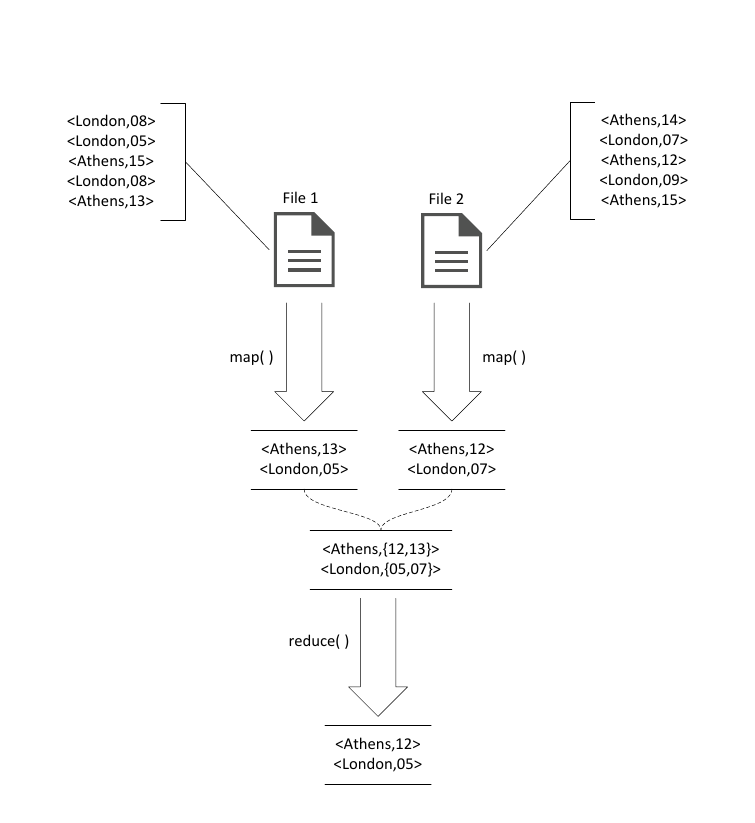}
\caption{An example of a MapReduce application with two parallel map tasks and one reduce task.}
\label{fig:MapReduce}
\end{figure}

A simple example of a MapReduce application with two parallel map tasks and one reduce task, is shown in Fig.~\ref{fig:MapReduce}. In the illustrated example, the overall minimum temperature recorded in London and Athens in a five-day period needs to be calculated for each city. It is assumed that the minimum temperature for each city was recorded daily in the form $\left<City, Minimum Temperature\right>$. The records are split into two files. Each file is processed in parallel by a map task. Each map task outputs the pairs that correspond to the minimum temperature for each city, according to the file that was processed. The results of the two map tasks are merged into two pairs (one for each city) in the form $\left<City, \left\{List Of Minimum Temperatures\right\}\right>$. The pairs are fed as input into the reduce task, which outputs the overall minimum temperature recorded in each city, over the said period. This parallel processing approach is more efficient than calculating the minimum temperature for each city in a serial fashion.      

An open source - and the most popular - implementation of the MapReduce programming model is the Apache Hadoop framework~\cite{Hadoop}, which adopts a master-slave architecture in order to exploit data locality. Specifically, the master node is responsible for scheduling the map tasks  of an application on the slave nodes, which contain chunks of the required input data. The reduce task is performed by the master node. When a slave node notifies the master node that it can accept a task, the master node scans the waiting tasks in queue to find the one that can achieve the best data locality. That is, the map task that its input data are located the nearest to the particular slave node is selected. However, due to the fact that Hadoop considers only one slave node at a time in order to schedule the map tasks, there are cases where it does not exploit data locality effectively. Furthermore, it cannot be employed for multi-cluster processing and for data-intensive applications that require more complex communication and processing patterns than those supported by the MapReduce paradigm.

\subsubsection{Other Approaches}
\label{subsubsec:otherApproaches}
In an attempt to tackle the aforementioned shortcomings of Hadoop and MapReduce, various approaches have been investigated in the literature. Among them, the \emph{delay scheduling} technique has been proposed, in order  to delay the scheduling of the waiting map tasks in case a slave node does not contain their input data, assuming that another slave node that contains the data will become available in a short period of time~\cite{Zaharia2010}. However, the drawback of this approach is that it wastes valuable time postponing the scheduling of the tasks, in an attempt to achieve better data locality, which is a goal that is not guaranteed. In order to overcome the single-cluster deployment restriction of the Hadoop framework, G-Hadoop has been proposed~\cite{Wang2013}. It is an extension of the traditional Hadoop framework that can schedule tasks across nodes of multiple clusters~\cite{Zhao2014}. For the scheduling of more complex data-intensive applications, various approaches have been proposed, utilizing the workflow scheduling paradigms described in Sect.~\ref{subsubsec:workflowScheduling}.

\subsection{Time Constraints}
\label{subsec:timeConstraints}
The most common QoS requirement that data-intensive applications may impose, is to finish execution within a strict time constraint. Such applications are regarded as \emph{real-time}, since they have a deadline that must be met~\cite{Stankovic1998}.

\subsubsection{Real-Time Applications}
\label{subsubsec:realTimeApplications}
\begin{figure}[t]
\centering
\includegraphics[scale=0.60]{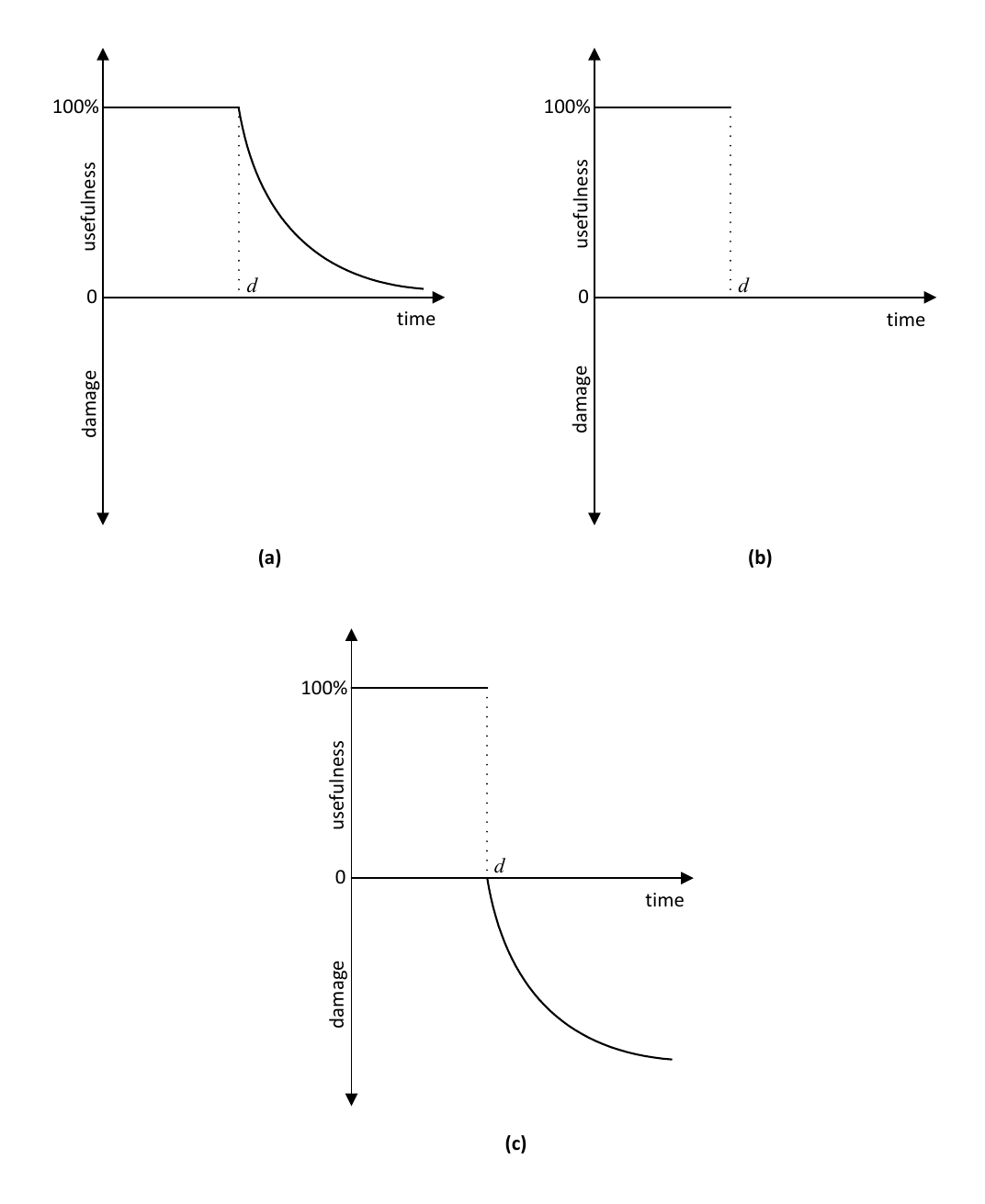}
\caption{The usefulness of the results of an application with a deadline $d$ over time, when $d$ is: (a) soft, (b) firm and (c) hard.}
\label{fig:allDeadlines}
\end{figure}

Depending on the severity of a missed deadline, real-time applications are classified into the following categories~\cite{Buttazzo2011}:
\begin{itemize}
\item \emph{Applications with soft deadlines}: in this case, the results of an application that missed its deadline still have some value, but their usefulness decreases with time (e.g. a user-system interaction application, where a delayed response to the user input is tolerated, degrading, however, the user experience as the delay increases).
\item \emph{Applications with firm deadlines}: in this case, the results will be useless, but this does not have any catastrophic consequences (e.g. a video streaming application, where a delayed video frame that arrives after the previous one on the user's terminal is discarded, since there is no value in playing it back). 
\item \emph{Applications with hard deadlines}: in this case, not only will the results be useless, but missing the application's deadline  will have catastrophic consequences. In this case, the damage caused by missing the deadline increases with time (e.g. a healthcare monitoring application, where a delayed analysis of patients data may cause loss of lives).
\end{itemize}
The impact of missing an application's deadline, as described above, is shown schematically in Fig.~\ref{fig:allDeadlines}. 

Two of the most widely used  policies for the scheduling of real-time data-intensive applications are the \emph{Earliest Deadline First (EDF)} and the \emph{Least Laxity First (LLF)} algorithms~\cite{Liu1973,Mok1983}. According to the EDF strategy, the component task  with the highest priority for execution is the one with the earliest deadline. On the other hand, according to the LLF policy, the task with the highest priority is the one with the minimum \emph{laxity}. The laxity of a task at a specific time instant, is defined as the difference between its deadline and its finish time. That is, it is the maximum amount of time that the particular task can delay its execution and still not miss its deadline.

A heuristic for the scheduling of real-time workflow applications in distributed systems, is the \emph{Least Space-Time First (LSTF)} policy~\cite{Cheng1997}, which takes into account both the precedence and the time constraints among the tasks. Specifically, according to this method, the task with the highest priority for scheduling is the one with the minimum value of the \emph{space-time} parameter. The space-time parameter of a task at a specific time instant, is defined as the difference between the deadline of the DAG and the level of the particular task. Even though this algorithm outperforms other scheduling policies, such as EDF, LLF and HLF described earlier, in the sense that it minimizes the maximum tardiness of the tasks, however, it exhibits poorer performance at guaranteeing deadlines. 

\subsubsection{Approximate Computations}
\label{subsubsec:approximateComputations}
\begin{figure}[h]
\centering
\includegraphics[scale=0.75]{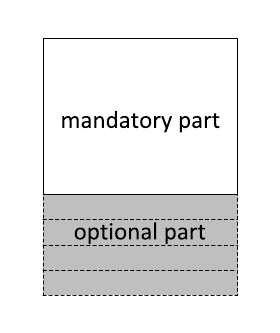}
\caption{A monotone task}
\label{fig:monotoneTask}
\end{figure}

Based on the observation that it is often more desirable for a real-time application to produce an approximate result by its deadline, than to produce a precise result late, the technique of \emph{approximate computations} has been proposed~\cite{Lin1987}. According to this method, a real-time application is allowed to return intermediate, approximate results of poorer, but still acceptable quality, when the deadline of the application cannot be met. Approximate computations can be utilized especially in the case of applications with \emph{monotone} component tasks, where the quality of a task's results is improved as more time is spent to produce them (e.g.  statistical estimation and video processing tasks). Each monotone task typically consists of a \emph{mandatory part}, followed by an \emph{optional part}. In order for a task to return an acceptable result, its mandatory part must be completed. The optional part refines the result produced by the mandatory part~\cite{Stavrinides2010, Stavrinides2011b}. A monotone task is illustrated in Fig.~\ref{fig:monotoneTask}.

Consequently, the approximate computations technique provides scheduling flexibility, by trading off precision for timeliness, since it allows the scheduler to terminate a task that has completed its mandatory part at any time, depending on the workload conditions of the system. For example, a video-on-demand server which streams video content to users over the Internet can benefit from this technique. The server may unexpectedly encounter network congestion, causing delays during the transmission of video content to users. Approximate computations can allow the system to reduce the quality of some video frames during a transmission, by omitting their optional enhancement layers and leaving only their base layer, so that the delivered video maintains an acceptable frame rate.

\subsection{Fault Tolerance}
\label{subsec:faultTolerance}
Fault tolerant scheduling in large-scale distributed systems, such as clouds,  is usually achieved through \emph{application-directed checkpointing}, which in contrast to system-directed checkpointing, is more practical, easier to implement and system-independent~\cite{Oldfield2007}. According to this approach, each application is responsible for checkpointing its own progress periodically, at regular intervals during its execution. In parallel data-intensive applications in particular, each component task periodically stores its state and intermediate data on persistent storage, creating a local checkpoint. The set of the local checkpoints (one from each task) that form a consistent application state, constitute a consistent global checkpoint. 

When a failure occurs, the application is rolled back and resumes execution from its last consistent global checkpoint. Checkpointing is a reactive failure management technique, where recovery measures are taken after the occurrence of a failure. As opposed to proactive failure management approaches, where prevention measures are taken before the occurrence of a failure (e.g. task migrations), reactive management is simpler to implement, since it does not require any complex failure prediction methods.

\subsection{Energy Efficiency}
\label{subsec:energyEfficiency}
There is a growing focus on  \emph{green computing} from both the academia and the industry, in an attempt to minimize the carbon footprint of data centers and increase the energy efficiency of applications. Typically, in most computing systems the processor consumes the greatest amount of energy compared to other components~\cite{Valentini2013, Stavrinides2017b}.  In embedded systems,  as well as in large-scale virtualized platforms such as the cloud, a technique that is frequently used in order to meet the energy constraints is the \emph{Dynamic Voltage and Frequency Scaling (DVFS)} method. This technique allows the dynamic adjustment of the supply voltage and operating frequency (i.e. speed) of a processor, based on the workload conditions, in an attempt to reduce the energy consumption of the processor~\cite{Kolodziej2012, Terzopoulos2014}. 

A heuristic frequently used with DVFS, is the \emph{slack reclamation} technique~\cite{Chen2006}. This method is based on the fact that the actual execution time of tasks is sometimes much shorter than their estimated worst case execution time. The difference between the actual and the worst case execution time of a task is called \emph{slack time}. At runtime, the scheduler tries to reclaim the slack time due to the early completion of a task, by selecting an unprocessed task to be executed at a slower processor speed via DVFS and thus save energy. 

An energy-efficient scheduling strategy for real-time BoT applications in the cloud utilizing DVFS, is the \emph{Cloud-Aware Energy-Efficient Scheduling (CAEES)} algorithm~\cite{Calheiros2014}. At each scheduling step, this method attempts to reduce the total energy consumption of the hosts, by selecting the most suitable virtual machine (VM) for the execution of each task, in an energy-wise manner. Specifically, the algorithm tries to schedule a task by examining specific criteria, starting from the best solution and gradually going to the worst solution: (a) the task is scheduled to a VM in use, without requiring an increase in its frequency, (b) the task is scheduled to a VM in use, but its operating frequency needs to be increased, (c) the task is scheduled to an idle VM, but there is at least one other VM on the same host that is not idle (i.e. the host is not idle) and (d) the task is scheduled to an idle VM on an idle host.

\section{Recent Novel Ideas and Research Trends}
\label{sec:novelStrategies}
In an attempt to provide even more effective scheduling solutions for data-intensive workloads in large-scale distributed systems, recent novel approaches have been proposed in the literature. As virtualization technologies evolve, a growing trend is the use of \emph{VM live migrations},  in order to better exploit data locality. Another prominent research trend is the utilization of approximate computations in combination with other techniques, in order to achieve better scheduling performance, in terms of timeliness, resilience against failures and energy conservation. For example, approximate computations can be combined with:
\begin{itemize}
\item bin packing techniques, in order to enhance timeliness,
\item checkpointing, in an attempt to improve fault tolerance and
\item DVFS, for better energy efficiency. 
\end{itemize}

\subsection{VM Live Migrations}
\label{subsec:VMliveMigrations}
In virtualized platforms, the VM live migration technique refers to the process of moving a running VM from one physical host to another, without downtime. That is, with no impact on the availability of the VM to the end-users and without interrupting the applications currently running on the VM. The memory, storage and network connectivity of the VM are transferred from the initial physical host to the destination host. Currently, the predominant use of VM live migrations, is to enhance energy efficiency and load balancing through server consolidation~\cite{Beloglazov2012}. 

However, the utilization of VM live migrations can also be used to better exploit data locality. Specifically, a virtualization approach has been proposed, where different VMs are used for each compute node and each storage node in the cloud~\cite{Sun2014}. In contrast to the  traditional approach where each compute and storage node are combined into one VM, this approach provides better flexibility and scalability, since compute nodes and storage nodes can be added or removed from the cloud independently. More importantly, according to this approach, a much lower live migration cost is incurred by migrating a compute node VM, compared to the traditional approach, where large volumes of data should be transferred to the destination host, since a VM would be both a compute and a storage node. In this framework, a data-aware scheduling method, \emph{DSFvH}, is employed, according to which live migrations of compute node VMs are performed, in order to place each compute node VM on the physical host that runs the storage node VM that contains the data required by the tasks executing on the compute node VM. This way, better exploitation of data locality is achieved.

 \subsection{Approximate Computations with Bin Packing}
 \label{subsec:ACBinPacking}
 The traditional \emph{bin packing} problem concerns the packing of a set of objects into a set of bins, using as few bins as possible~\cite{Coffman2013}. The most commonly used bin packing techniques are: (a) \emph{First Fit (FF)}, where the object is placed into the first bin where it fits, (b) \emph{Best Fit (BF)}, where  the object is placed into the bin where it fits and leaves the minimum unused space possible and (c) \emph{Worst Fit (WF)}, where the object is placed into the bin where it fits and leaves the maximum unused space possible. 
 
 In an attempt to improve the timeliness of real-time workflow applications  in a heterogeneous distributed system, a novel list scheduling heuristic has been proposed, which utilizes schedule gaps with a technique that combines approximate computations with the FF, BF and WF bin packing policies~\cite{Stavrinides2012, Stavrinides2015}.  Another characteristic of the proposed approach, is that it takes into account the effects of error propagation among the tasks of an application in case of partially completed tasks.  The task prioritization is based on EDF. Once a task is selected by the scheduler, it is allocated to the processor that can provide it with the earliest estimated start time. In order to calculate the estimated start time of the task on the particular processor, schedule gaps are exploited with a technique that allows only a fraction of the task to be inserted into an idle time slot. The fraction of the task to be inserted into a schedule gap must be at least equal to the mandatory part of the task. Moreover, its potential output error must not exceed the input error limit of its child tasks. 
 
\begin{figure}[!t]
\centering
\includegraphics[scale=1]{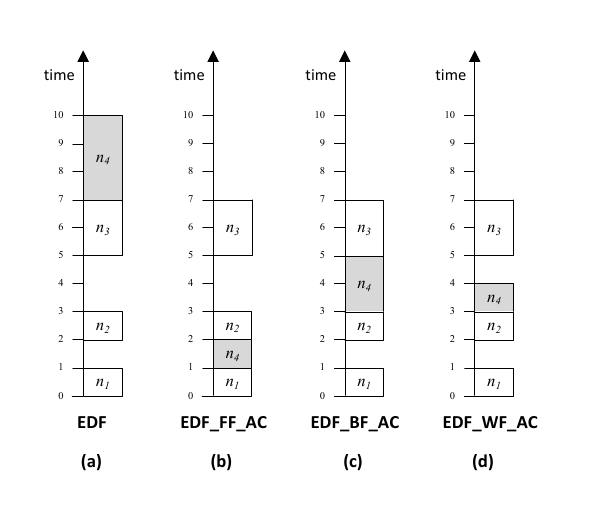}
\caption{An example of scheduling tasks with the strategies described in Sect.~\ref{subsec:ACBinPacking}. A task $n_{4}$ is scheduled  according to one of the policies: (a) EDF (baseline algorithm), (b) EDF\_FF\_AC, (c) EDF\_BF\_AC and (d) EDF\_WF\_AC. The parameters of the tasks used in the example are shown in Table~\ref{tab:BinPackingAC}.}
\label{fig:BinPackingAC}
\end{figure}

 The placement of the partial task into a schedule gap is performed using a modified version of either the FF, BF or WF bin packing policy:
 \begin{itemize}
 \item \emph{First Fit with Approximate Computations (FF\_AC)}: the task is placed into the first schedule gap where at least its minimum possible computational cost fits.
 \item \emph{Best Fit with Approximate Computations (BF\_AC)}: the task is placed into  the schedule gap where its maximum possible computational cost fits, leaving the minimum unused time possible.
 \item \emph{Worst Fit with Approximate Computations (WF\_AC)}: the task is placed into the schedule gap where its minimum possible computational cost fits, leaving the maximum unused time possible.
 \end{itemize}
 In contrast to this approach, the other list scheduling heuristics presented earlier, ISH and  HEFT, essentially use FF in order to utilize idle time slots. More importantly, with the incorporation of approximate computations, this approach is more flexible, allowing only a fraction of  a task to be inserted into a schedule gap when the task does not completely fit into it. An example of scheduling tasks with the proposed heuristics (EDF\_FF\_AC, EDF\_BF\_AC and EDF\_WF\_AC), compared to the baseline EDF policy, is illustrated in Fig.~\ref{fig:BinPackingAC}. The parameters of the tasks used in the example are shown in Table~\ref{tab:BinPackingAC}.

\begin{table}
\caption{The parameters of the tasks used in the example of Fig.~\ref{fig:BinPackingAC}. For each task, $d$ is its deadline, $t_{data}$ is the time at which its required input data will be available, $c$ is its computational cost and $c_{min}$ is its minimum computational cost when approximate computations are utilized.}
\label{tab:BinPackingAC}
\begin{tabular}{p{2.2cm}p{2.2cm}p{2.2cm}p{2.2cm}p{2.2cm}}
\hline\noalign{\smallskip}
$Task$ & $d$ & $t_{data}$ & $c$  & $c_{min}$\\
\noalign{\smallskip}\svhline\noalign{\smallskip}
$n_{1}$  & $2$ & $0$ & $1$ & $1$\\
$n_{2}$ & $4$ & $2$ & $1$ & $1$\\
$n_{3}$ & $9$ & $5$ & $2$ & $1$\\
$n_{4}$ & $10$ & $1$ & $3$ & $1$\\
\noalign{\smallskip}\hline\noalign{\smallskip}
\end{tabular}
\end{table}
 
\subsection{Approximate Computations with Checkpointing}
\label{subsec:ACCheckpointing}
In an attempt to improve resilience against transient software failures in a SaaS cloud, where real-time fine-grained parallel applications are scheduled and executed, the approximate computations technique has been combined with application-directed checkpointing~\cite{Stavrinides2008, Stavrinides2009, Stavrinides2016b}. Specifically, gang scheduling is employed, where the prioritization of the component tasks is according to the EDF policy. In addition to application-directed checkpointing, fault tolerance is enhanced by the use of approximate computations in either a restricted manner or a more holistic approach. In the first case, an application may provide approximate results when it has completed its parallel mandatory part and (a) its deadline is reached, (b) a failure occurred and its last generated checkpoint stored results corresponding to computational work greater than or equal to its mandatory part or (c) another notified application must start execution immediately (i.e. there is time to execute only the mandatory part of the other application before its deadline). According to the second approach, all applications are scheduled to complete only their mandatory part. That is, in this case all applications give approximate results.

 \subsection{Approximate Computations with DVFS}
\label{subsec:ACDVFS}  
In order to enhance energy efficiency, a heuristic that combines approximate computations with DVFS has been proposed, for the scheduling of periodic real-time tasks~\cite{Mizotani2015}. According to this approach, the tasks are scheduled according to the \emph{Mandatory-First Earliest Deadline (MFED)} policy, while the supply voltage and processor frequency are scaled according to the \emph{Cycle-Conserving Real-Time DVFS (CC-RT-DVFS)} technique. MFED  is a policy according to which the mandatory parts of the tasks have always higher priority than the optional parts. The mandatory part with the earliest deadline has the highest priority for execution. CC-RT-DVFS is essentially a dynamic slack reclamation technique, which utilizes the slack time that occurs due to the early completion of a mandatory part, for the scheduling of the optional part of the task at a lower processor speed, utilizing DVFS.  Thus, in this strategy there is a trade-off not only between result precision and timeliness, but also between result precision and energy savings.

\section{Conclusions}
\label{sec:conclusions}
 In this chapter, a classification of data-intensive workloads was proposed and an overview of the most commonly used heuristics for their scheduling in large-scale distributed systems was given.  Major challenges of data-intensive applications were covered, such as data locality awareness, timeliness, resilience against failures and energy efficiency. Furthermore, recent novel ideas and research trends were presented. 
 
 Scheduling data-intensive workloads in large-scale distributed systems remains an active research area, with many open challenges. With the explosive growth of big data, workloads tend to get more complex and computationally demanding. Consequently, more effective scheduling heuristics must be employed. In addition to the data locality awareness, timeliness, fault tolerance and energy efficiency objectives, security is drawing an ever-increasing interest from both the industry and the research community. Hence, efforts towards this direction are expected to be intensified in the near future. 
 
 \begin{acknowledgement}
 The second author of this chapter, Helen D. Karatza, has been invited as a trainer to the cHiPSet Training School 2016 \emph{``New Trends in Modeling and Simulation in HPC Systems''}, held in Bucharest, Romania, 21-23 September 2016, and has been supported by the IC1406 Horizon 2020 grant.
 \end{acknowledgement}

\bibliography{GStavrinidesHKaratza.bib}

\end{document}